# High-efficiency WSe$_2$ photovoltaics enabled by ultra-clean van der Waals contacts

Kamal Kumar Paul[1]*, Cullen Chosy[2,3], Soumya Sarkar[1], Zhuangnan Li[1], Han Yan[1], Ye Wang[1], Leyi Loh[1], Lixin Liu[1], Hu Young Jeong[4], Samuel D. Stranks[2,3], Yan Wang[1], and Manish Chhowalla[1]*

[1]Department of Materials Science & Metallurgy, University of Cambridge, United Kingdom

[2]Department of Chemical Engineering and Biotechnology, University of Cambridge, Cambridge, United Kingdom

[3]Cavendish Laboratory, University of Cambridge, Cambridge, United Kingdom

[4]UNIST Central Research Facilities (UCRF) and School of Materials Science and Engineering, UNIST, Ulsan, South Korea

*kkp26@cam.ac.uk, mc209@cam.ac.uk

**Abstract.** Layered transition metal dichalcogenide semiconductors are interesting for photovoltaics owing to their high solar absorbance and efficient carrier diffusion. Tungsten diselenide (WSe$_2$), in particular, has emerged as a promising solar cell absorber. However, defective metal-semiconductor interfaces have restricted the power conversion efficiency (PCE) to ~ 6 %. Here we report WSe$_2$ photovoltaics with a record-high PCE of ~11% enabled by ultra-clean indium/gold (In/Au) van der Waals (vdW) contacts. Using grid-patterned top vdW electrodes, we demonstrate near-ideal diodes with a record-high on/off ratio of $1.0\times10^9$. Open-circuit voltage ($V_{OC}$) of 571 ± 9 mV, record-high short-circuit current density ($J_{SC}$) of 27.19 ± 0.45 mA cm$^{-2}$ – approaching the theoretical limit (34.5 mA cm$^{-2}$) – and fill factor of 69.2 ± 0.7% resulting in PCE of 10.8 ± 0.2% under 1-Sun illumination on large active area (~0.13×0.13 mm$^2$) devices have been realised. The excellent device performance is consistent with the high external quantum efficiency (up to ~93%) measured across a broad spectral range of 500–830 nm. Our results suggest that ultra-clean vdW contacts on WSe$_2$ enable high-efficiency photovoltaics and form the foundation for further optimisation.

**Introduction.** Layered transition metal dichalcogenides (TMDs) such as molybdenum disulfide (MoS$_2$), tungsten disulfide (WS$_2$), and tungsten diselenide (WSe$_2$) have emerged as efficient absorbers for thin solar cells.[1–3] Multilayer TMDs with a bandgap of 1–1.3 eV are ideal candidates for high-efficiency single-junction solar cells.[4,5] Strong light-matter interactions enable TMD semiconductors to achieve an order of magnitude higher absorption than GaAs and Si.[5–7] TMDs with broadband near-unity absorption possess self-terminating bonds, which minimize surface defects – leading to low surface recombination and high internal quantum efficiency (IQE).[8–10] A modified balance model indicates that a single-absorber TMD solar cell can achieve a power conversion efficiency (PCE) of up to 27%, assuming a perfect external quantum efficiency (EQE) and ideal carrier collection.[5]

Photovoltaic performance of WSe$_2$ solar cells remains in the early stages with a PCE of around 6 – 7%[11,12] – albeit for micron-scale (0.02×0.02 mm$^2$) devices. Despite recent advancements, the poor metal-semiconductor contacts and insufficient channel depletion[13] have hindered efficient photocarrier separation and collection in TMD-based solar cells.[14–16] It is well known that interfaces between metal contacts and TMD semiconductors are typically defective – resulting in midgap states[16], high contact resistance, and Fermi level pinning.[17] Thus, the Schottky-Mott rule for Schottky barrier height does not apply to metal-TMD interfaces – resulting in low open-circuit voltages.[18–20] Further, midgap states at defective interfaces act as recombination centres, leading to significant photocarrier trapping.[16] Several strategies, such as metal and graphene transferred contacts to reduce interface defects, have led to enhancement in the performance of TMD photovoltaics.[2,16,20] However, even in relatively clean transferred contacts, adsorbed contaminants at the interface can cause Fermi level pinning and result in high contact resistance,[17] reducing the rectification ratio.[1,11,21] A key challenge in the TMD solar cell community is identifying suitable top contacts with high optical transmittance for efficient sunlight illumination.

In this work, we demonstrate multilayer $WSe_2$-based single-absorber solar cells with ultra-clean vdW contacts. The photovoltaic devices consist of platinum (Pt) as the bottom electrode and indium/gold (In/Au) alloy vdW contacts pioneered in our group[22] as the top contacts. The devices with vdW contacts show near-ideal diode behaviour with a record-high on/off ratio of $1.0 \times 10^9$ – indicating negligibly low interface traps and leakage current. The clean interfaces and high-quality semiconductor enable high external quantum efficiency (EQE) across a broad spectral range (500–830 nm), contributing to a record PCE up to ~11%. Further, we describe a grid-patterned top electrode design with ultra-clean vdW interfaces that minimizes reflection losses in the active area while maintaining efficient charge extraction. A PCE of ~10.8 ± 0.2% has been achieved on large active area devices (~0.13×0.13 mm²), with an open-circuit voltage ($V_{OC}$) of 571 ± 9 mV, fill factor (FF) of 69.2 ± 0.7% and record-high short-circuit current density ($J_{SC}$) of 27.19 ± 0.45 mA cm$^{-2}$.

**Results and discussion.**

To fabricate the devices, $WSe_2$ flakes were mechanically exfoliated and transferred onto high work function Pt electrodes to achieve ohmic hole contacts (see Methods for details of $WSe_2$ transfer technique). Grid-patterned top electrodes with low work function were directly deposited on the $WSe_2$ [see Fig. 1a(i)]. The grid size was carefully optimized to balance effective optical excitation of the TMD absorber with efficient photocarrier collection, thereby maximizing the overall solar cell efficiency. The simple single-absorber photovoltaic device structure is shown in Fig. 1a(i). In the devices, excitons are generated under illumination in $WSe_2$ and dissociated by the built-in potential induced by the difference in work function between the top and bottom electrode metals. For the top electrodes, we used vdW and non-vdW contacts (see Methods for metal electrode deposition conditions) for comparison.

In conventionally fabricated Schottky diodes, the evaporation of metal electrodes damages $WSe_2$ through kinetic energy transfer or chemical reactions, leading to non-vdW interfaces with defect states and Fermi-level pinning [Fig. 1a (ii, iv)].[17] In our previous studies, we have shown that it is possible to achieve defect-free vdW interface with In/Au alloy contacts and TMDs [Fig. 1(iii,v)].[22] The atomic resolution high angle annular dark field-scanning transmission electron microscopy (HAADF-STEM) images of defective and ultra-clean In/Au vdW contacts are shown in Fig. 1b,c, respectively. In Fig. 1b, intermixing of metal electrode with $WSe_2$ – indicating damage and presence of structural defects – can be clearly observed.

In contrast, Fig. 1c shows that the WSe$_2$ layer adjacent to the In/Au contact is pristine and a clear vdW gap can be seen between the two – indicating ultra-clean interface. Low-magnification HAADF-STEM images over a large area further illustrate the contrasting interfacial quality of defective (Ti/Au or Al/Au) and ultra-clean (In/Au) vdW contacts with WSe$_2$, highlighting widespread damage in the former and uniform, defect-free vdW interfaces in the latter (Fig. S1 Supporting Information). Interface defects create mid-gap states that trap electrons – reducing the $V_{OC}$ and current density of the device [Fig.1a (iv)]. The clean interface with vdW gap facilitates electrons to flow in forward direction and suppresses current in the reverse direction so that photovoltaic diodes with significantly improved performance can be achieved.

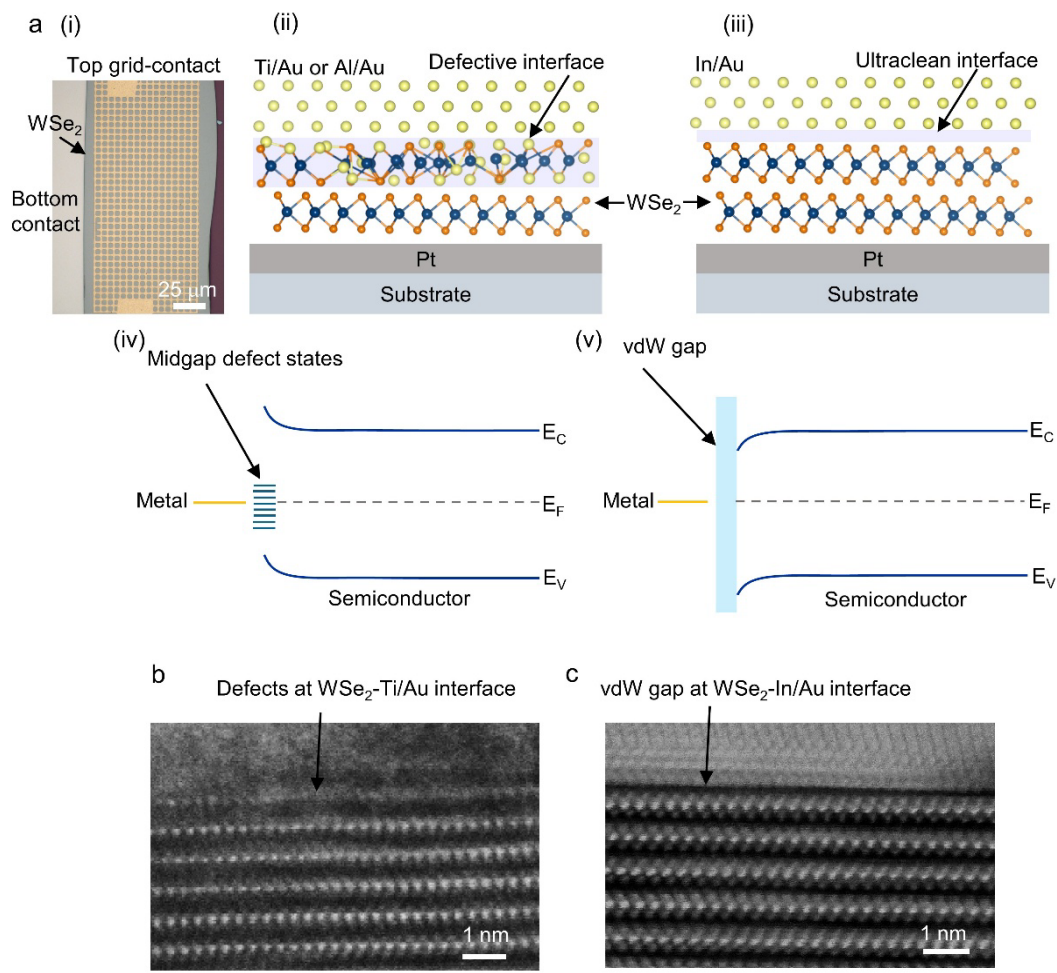

**Fig. 1: Vertical solar cells with defective and ultra-clean vdW interfaces. a,** Vertical diodes with high work function bottom contacts (Pt, Pd) and low work function grid-patterned top contacts (Ti/Au, Al/Au, In/Au). (i), Optical micrograph of a representative device. (ii), Schematic side-view of device with Ti/Au or Al/Au top contacts, showing interface defects at the top metal-WSe$_2$ interface. (iii) Interface with In/Au shows a vdW gap for a pristine and ultra-clean contact. Energy level diagram for (iv) defective interface showing midgap defect states that pin the Fermi level, resulting in a higher Schottky barrier for electron transport and reduced current density, and (v) ultra-clean WSe$_2$-In/Au contact with a vdW gap, enabling a lower Schottky barrier and enhanced charge transport. **b,** Atomic-resolution STEM image showing damage/defects at the WSe$_2$-Ti/Au interface. **c,** Ultra-clean WSe$_2$-In/Au interface with a clearly visible vdW gap.

Diode characteristics of devices with vdW (In/Au) and defective (Al/Au, Ti/Au) top contacts are shown in Fig. 2a. It can be seen that the defective contact diodes exhibit significantly higher leakage currents in reverse bias compared to the vdW contact diodes. By contrast, In/Au top contact diodes show significantly lower reverse leakage current of $2.45\times10^{-12}$ A and higher rectification ratio of $1.0\times10^9$. The reduced forward current and early saturation observed in defective contact diodes are attributed to higher Schottky barrier heights and increased series resistance. Mid-gap defect states at the metal–semiconductor interface contribute to elevated reverse leakage through trap-assisted conduction. This increased series resistance stems from Fermi level pinning and the presence of defect states, which hinder efficient charge injection and introduce resistive barriers. In contrast, ultra-clean vdW contact diodes show enhanced forward current due to a pristine, defect-free interface that prevents Fermi level pinning and enables a lower Schottky barrier. This leads to efficient carrier injection and reduced series resistance. Under reverse bias, the lack of mid-gap defect states effectively blocks leakage pathways, yielding a substantially lower reverse current.

The quality of a diode can be assessed by the ideality factor (1.0 for a perfect diode) determined by fitting the Shockley diode equation to dark current measurements (see Supporting Information). We obtain an ideality factor of 1.1 for vdW contacts devices, which is among the lowest measured for diodes based on TMDs and two-dimensional (2D) semiconductors (Fig. 2b). I-V characteristics in dark for several different Pt-WSe$_2$-In/Au devices are shown in Fig. S2, Supporting Information. The near-unity ideality factor indicates that the diodes with In/Au contacts are mostly free from interface trap states, Fermi level pinning, and negligible current in reverse bias. Further, the on/off ratio of vdW contact diodes was found to be $> 1.0\times10^9$ – the highest measured to date for this class of materials (Fig. 2b).

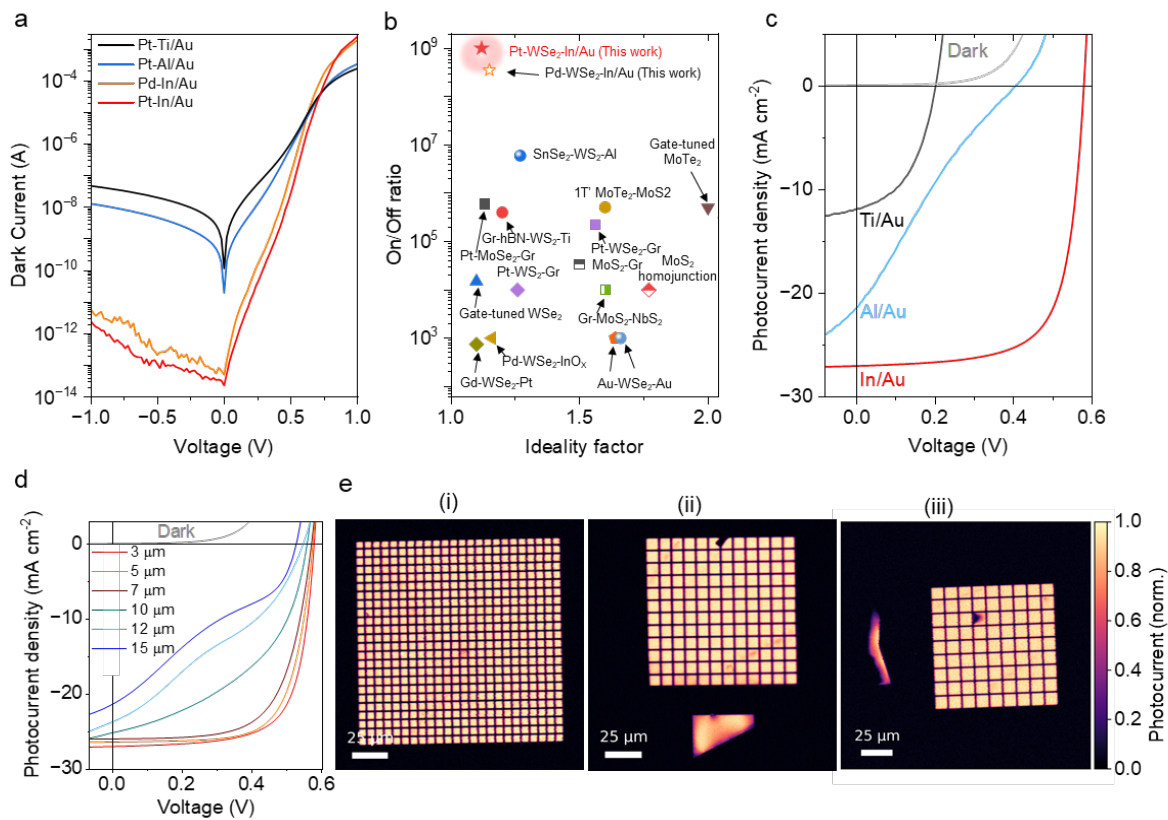

**Fig. 2: Electrical and photovoltaic properties (under 1-Sun) of diodes with defective and ultra-clean vdW contacts. a,** I-V curves of diodes with defective and vdW contacts in dark. The ultra-clean vdW contact devices show lower leakage currents at negative biases and better ideality factor. **b,** Performance of the ultra-clean vdW contact devices and other reported diodes based on similar materials.[11,21,23–30] **c,** I-V curves in dark and under 1-Sun illumination of defective and ultra-clean vdW top contact diodes with Pt bottom contact. In/Au (red curve) contact shows significantly improved performance than Ti/Au (black curve) and Al/Au (blue curve) contacts. **d,** I-V curves for Pt-WSe$_2$-In/Au devices with different grid sizes (3 × 3 – 15 × 15 μm$^2$), showing significant degradation in performance as grid size increases beyond 7 × 7 μm$^2$. **e,** Photocurrent maps of Pt-WSe$_2$-In/Au devices with grid sizes of (i) 5 × 5 μm$^2$, (ii) 7 × 7 μm$^2$, and (iii) 10 × 10 μm$^2$, respectively.

The near ideal diode properties are translated into promising photovoltaic performance. Devices with different top contacts were investigated under 1-Sun AM 1.5G illumination. Typical current-voltage (I-V) curves of defective and ultra-clean vdW top contact devices with identical Pt bottom contacts and WSe$_2$ of similar thicknesses are shown in Fig. 2c. Devices with Al/Au and Ti/Au top contacts show low current densities, fill factor, and open-circuit voltages due to defective interfaces.[19–21] I-V curves of devices with bottom electrode of different work function metals and In/Au top contacts are shown in Fig. S3, Supporting Information, which show that the open-circuit voltage increases as the work function of the bottom electrode increases – as expected. The work function of bulk Pt crystals is ~5.6 eV[31], but is much lower (~ 5.0 eV) for evaporated polycrystalline thin film Pt[32]. Therefore, the work function difference between the bottom (Pt or Pd ~ 5.0 eV) and top (In/Au ~ 4.1 eV)[33] contacts is about 0.9 eV, which is reasonably close to the open-circuit voltage of ~ 0.6 V achieved in our ultra-clean vdWs contact Pt-WSe$_2$-In/Au and Pd-WSe$_2$-In/Au devices.

To optimize the charge collection efficiency, we examined Pt-WSe$_2$-In/Au devices with varying grid electrode sizes ranging from 3 × 3 to 15 × 15 μm$^2$. The I–V characteristics of these devices, fabricated with comparable WSe$_2$ thicknesses but different top contact grid sizes, are presented in Fig. 2d (WSe$_2$ thickness profiles are provided in Fig. S4 and S5, Supporting Information). For grid sizes up to 7 × 7 μm$^2$, the I–V curves display similar behaviour, featuring high photocurrent densities and well-defined diode-like characteristics, although there is a marginal decrease in photocurrent density with increasing grid size. However, for grid size beyond 7 × 7 μm$^2$, a sharp reduction in photocurrent density is observed (Fig. S6, Supporting Information), along with a noticeable decrease in the slope of the I–V curves near V$_{OC}$, indicating increased series resistance and reduced charge collection efficiency. These trends are consistent with the dark I–V characteristics (Fig. S7, Supporting Information), where grid size of ≤7 × 7 μm$^2$ shows high forward saturation currents and low series resistances (76–127 Ω). In contrast, devices with larger grid sizes exhibit reduced forward saturation currents and significantly increased series resistances, ranging from 741 Ω to 7.7 × 10$^3$ Ω, further confirming the importance of grid size optimization for superior device performance. Thus, the poor photovoltaic performance with grid size larger than 7 × 7 μm$^2$ can be attributed to high series resistance, originating from the limited top metal contact area.

Photocurrent mapping [Fig. 2e (i-iii)], corresponding to the devices presented in Fig. S8, Supporting Information, further corroborates these observations. Photocurrent maps of Pt-WSe$_2$-In/Au devices using 405 nm laser excitation reveal that photocurrent is uniformly collected over the square optical windows on WSe$_2$, with negligible contribution from areas shaded by the metal grid. It is important to note that the top metal grid is precisely aligned with the bottom electrode. To eliminate uncertainty in active area calculation, WSe$_2$ flakes outside the metal grid are further coated with In/Au to suppress any unwanted photocurrent contributions. As the bottom electrode only extends over the grid area, this increased top electrode area should not contribute to an increased dark current. Photocurrent outside the grid area is only observed in those regions where the WSe$_2$ layer bridges the bottom contact and is intentionally left unmasked to prevent short-circuiting. Both the optical windows and unmasked regions are included in calculating the active area, while the gridlines and peripheral regions shaded by the In/Au electrode are excluded. Devices with 5 × 5 μm$^2$, 7 × 7 μm$^2$ and 10 × 10 μm$^2$ grids exhibit high and spatially uniform photocurrent, demonstrating that photogenerated carriers are effectively collected. This effect is attributed to the long lateral carrier diffusion length in WSe$_2$.[34] However, high series resistance beyond 7 × 7 μm$^2$ grids associated with the limited top metal contact area significantly reduces carrier extraction efficiency, leading to substantial recombination losses before the carriers can reach the electrodes. Consequently, as shown in Fig. 3a and 3b, both the fill factor and the power conversion efficiency (PCE) drop sharply when the grid size exceeds 7 × 7 μm$^2$.

We also studied the influence of vertical diffusion of photocarriers by fabricating Pt-WSe$_2$-In/Au devices with different WSe$_2$ thicknesses. The variation of the photocurrent density as a function of the WSe$_2$ thickness is shown in Fig. 3c. Fitting the data with $J = J_0 - J_1 e^{-x/\lambda}$ (where $J$, $J_0$, and $J_1$ are current density as a function of depth, $x$, steady-state dark current density, and maximum current density, respectively), the characteristic length parameter for carrier diffusion ($\lambda$) was estimated to be ~201 nm. The PCE corresponding to the photocurrent of devices in Fig. 3c as a function of WSe$_2$ film thickness is plotted in Fig. 3d. It can be seen that the PCE increases with the WSe$_2$ film thickness up to ~ 200 nm, above which the PCE decreases. This is because of low absorption in WSe$_2$ below the diffusion length of ~ 200 nm, which leads to low J$_{SC}$ and, thus, reduced PCE. On the other hand, WSe$_2$ films thicker than the diffusion length suffer from recombination losses, as reflected in the low fill factor (Fig. S9, Supporting Information), leading to reduced PCE.

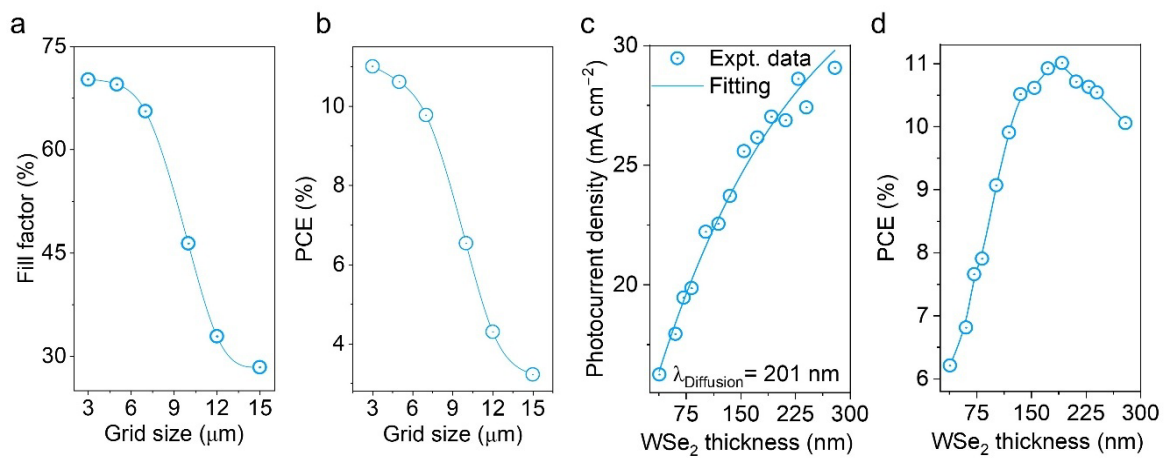

**Fig. 3: Photovoltaic parameters versus metal grid size and WSe$_2$ film thickness. a-b,** Fill factor (FF) and PCE, respectively, as a function of grid size. Both show a sharp decline for grid sizes larger than 7 × 7 μm$^2$. **c,** Variation of short-circuit current density (J$_{SC}$) in Pt-WSe$_2$-In/Au devices with different WSe$_2$ film thicknesses. The characteristic diffusion length (λ$_{Diffusion}$) is extracted by fitting with the diffusion length equation. **d,** Dependence of power conversion efficiency (PCE) on thickness of WSe$_2$ films.

We have made many devices to investigate the consistency in performance of the WSe$_2$ photovoltaics with vdW contacts. I-V characteristics of seven typical devices (3 × 3 μm$^2$ and 5 × 5 μm$^2$) are shown in Fig. 4a. The inset displays the average values of key photovoltaic parameters along with associated errors, which account for device-to-device variability and instrumental measurement errors. The I-V characteristics of one batch of devices are shown in Fig. S10a, Supporting Information. The variability in device performance is attributed to different thicknesses of WSe$_2$. The statistical distributions of the key photovoltaic parameters such as the open-circuit voltage (V$_{OC}$), short-circuit current density (J$_{SC}$), fill factor (FF), and the power conversion efficiency (PCE) are shown in Fig. S10b, Supporting Information. A record-high PCE of 10.8 ± 0.2% is achieved in large-area ~0.13×0.13 mm$^2$ solar cells. The photovoltaic parameters of various devices, along with their corresponding average values, are summarized in Table S1, Supporting Information. The maximum V$_{OC}$ and FF recorded are 580 mV and 70.2%, with their corresponding average values of 571 ± 9 mV (overall average: 564 ± 12 mV) and 69.2 ± 0.7% (overall average: 67.8 ± 2.4%), respectively. A particularly important result is that our ultra-clean vdW contact solar cells achieve a maximum J$_{SC}$ of 28.14 mA cm$^{-2}$ with an average value of 27.19 ± 0.45 mA cm$^{-2}$ (overall average: 27.23 ± 0.55 mA cm$^{-2}$), setting a new benchmark that further approaches the limit for J$_{SC}$ (34.5 mA cm$^{-2}$) in WSe$_2$-based single junction solar cells.[35]

A large-area Pt-WSe$_2$-In/Au device, fabricated by integrating multiple flakes with an active area of 0.32 × 0.32 mm$^2$ was also fabricated. It showed excellent photovoltaic performance, achieving an overall PCE of 8.2% (see Fig. S11, Supporting Information). Hysteresis is negligible in the I-V characteristics of Pt-WSe$_2$-In/Au devices when sweeping the bias voltage forward and backward as shown in Fig. S12, Supporting Information. Long-term operational stability of the hBN-encapsulated devices using continuous maximum power point (MPP) tracking under ambient conditions and 1-sun illumination (see Methods) is demonstrated in Fig. S13, Supporting Information. The device maintains a stable PCE with negligible degradation even after 30 days, demonstrating the robustness of TMD solar cells.

Fig. 4b shows the EQE spectrum as a function of wavelength (left Y-axis) and the integrated J$_{SC}$ extracted from the EQE spectrum (right Y-axis). EQE quantifies the fraction of photons incident on the solar cell absorber successfully converted into charge carriers that contribute to the current. An ideal solar cell converts all the incident photons into current. Our ultra-clean vdW contact solar cell demonstrates excellent quantum efficiency (up to ~93%) over a broad spectral range of 500-830 nm. Integrated J$_{SC}$ is estimated as 25.93 mA cm$^{-2}$ with an error margin of ~4%. This value is in close agreement with the corresponding value obtained from the I-V curve (Fig. 4a) under 1-Sun. To further evaluate the influence of contact geometry, EQE spectra were measured for devices with grid sizes of 7 × 7 μm$^2$ and 15 × 15 μm$^2$, as shown in Fig. S14, Supporting Information. The corresponding integrated $J_{SC}$ are 24.6 mA cm$^{-2}$ and 20.3 mA cm$^{-2}$, respectively, each showing a ~4% deviation from values extracted from the 1-Sun I–V curves (Fig. 2d). The observed decrease in photocurrent with increasing grid size is primarily attributed to a rise in series resistance beyond the 7 μm grid, which hinders efficient charge extraction. Additionally, larger grid sizes introduce longer lateral carrier transport paths and increased recombination losses, further contributing to the reduced collection efficiency. Photocurrent maps collected under a calibrated excitation intensity can be converted to show EQE (Fig 4c), agreeing with the corresponding macroscopic value at 405 nm (Fig. 4b). Further measurement of local reflectance losses allows for calculation of IQE, showing peak values approaching unity (~98%), as shown in Fig. S15, Supporting Information. This indicates that internal losses are exceptionally low as nearly all absorbed photons are efficiently converted into charge carriers that contribute to the photocurrent.

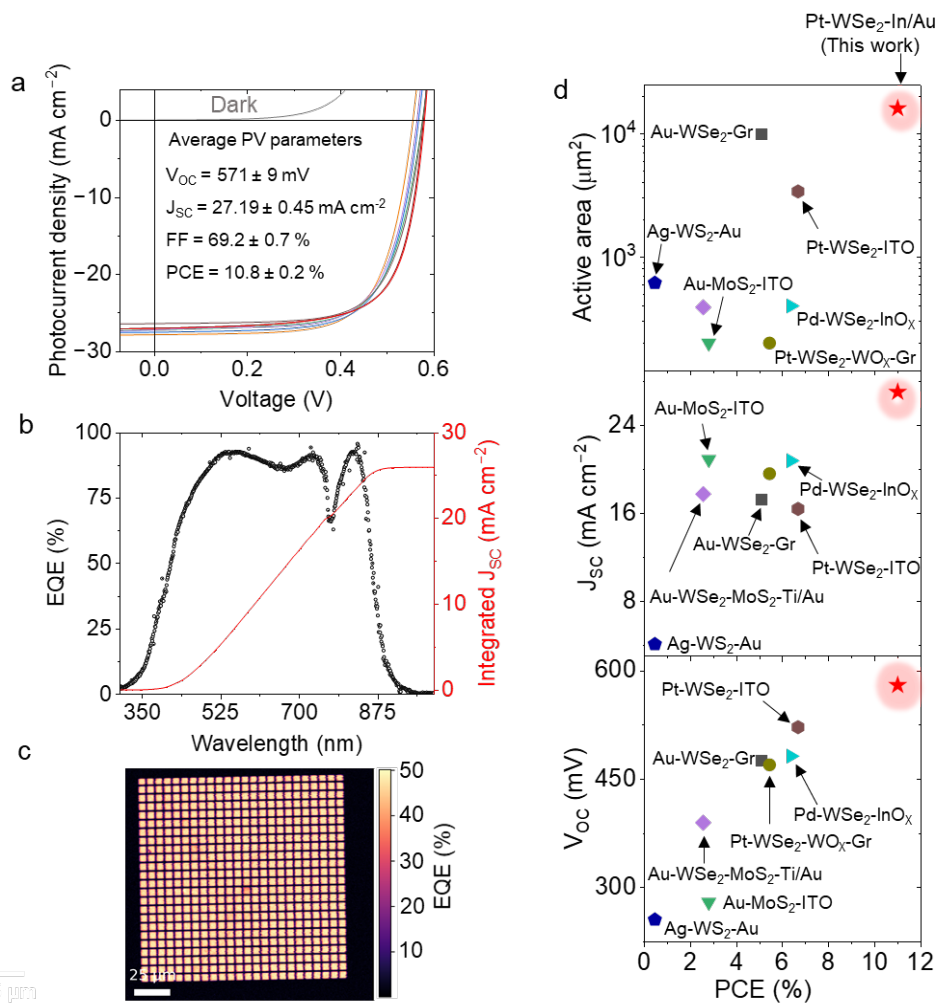

**Fig. 4: Pt-WSe$_2$-In/Au photovoltaic devices under 1-Sun illumination (AM 1.5G) and benchmarking. a,** I-V curves measured on five Pt-WSe$_2$-In/Au devices. Inset shows average photovoltaic parameters with errors. The error analysis accounts for device-to-device variability, and instrumental measurement errors. **b,** Measured EQE spectrum (left Y-axis) along with the integrated $J_{SC}$ obtained from the EQE curve (right Y-axis). **c,** EQE map of the device with a grid size of 5 × 5 μm² under 405 nm laser excitation, shows good agreement with the macroscopic EQE at 405 nm in panel b. **d,** Benchmark of $V_{OC}$, $J_{SC}$, and active area versus

PCE in Pt-WSe$_2$-In/Au device compared to different TMD photovoltaic devices under 1-Sun illumination. Star symbols represent our best Pt-WSe$_2$-In/Au devices measured in different batches.[1–3,11,12,36–38] See Table S2, Supporting Information for details.

Fig. 4d compares the different photovoltaic parameters, including V$_{OC}$ (bottom panel), J$_{SC}$ (middle panel), and active area (top panel) of our ultra-clean vdW contact devices with previously reported vertical TMD photovoltaic devices under 1-Sun illumination. Details are shown in Table S2, Supporting Information. A comprehensive performance comparison of 2D vdW materials-based solar cells, including lateral[39–41], gated[40,42,43], doped[1,38], and heterostructure[37,44,45] designs under 1-Sun illumination demonstrate that the PCE has previously been limited to a maximum of 6.7%.[12]

In summary, our single absorber WSe$_2$ solar cells establish a new benchmark with a high on/off ratio of $1.0\times10^9$, V$_{OC}$ of 580 mV, J$_{SC}$ of 27.03 mA cm$^{-2}$, FF of 70.2%, and high PCE of ~11% enabled by defect-free vdW contacts. The ultra-clean metal-semiconductor interfaces promote efficient carrier collection, thus approaching the theoretical J$_{SC}$. The grid-patterned top electrode design with optical windows on WSe$_2$ eliminates the need for alternative transparent electrodes. Our work opens up opportunities for further improvement in V$_{OC}$ and fill factor (FF). Our industry-compatible and simple fabrication approach will enable the realization of high-performance photovoltaics.

**Methods**

**Device fabrication**

High-quality bulk WSe$_2$ crystals (2D Semiconductors, USA) with various thicknesses (~38 − 300 nm) were mechanically exfoliated on UV-treated PDMS stamps. Prior to exfoliation, the PDMS was exposed to ultraviolet light for 20 minutes to minimize residues on the WSe$_2$ surface. This reduction in PDMS residues is confirmed by atomic force microscopy analysis (Fig. S16, Supporting Information). Thereafter, they were transferred to the pre-patterned bottom electrodes (Au, Pd, and Pt) with a thickness of 25 nm. The bottom electrodes are designed by a shadow mask using photolithography, followed by metal deposition and lift-off process. The transferred WSe$_2$ flakes on high work function metal electrodes were then annealed in an Ar/H$_2$ atmosphere at 250 °C for 1 hour. The samples were then coated with MMA/PMMA and patterned using electron-beam lithography. Before metal electrode deposition, the electron beam evaporation system was evacuated to a base pressure of <10$^7$ Torr. Subsequently, a 30 nm-thick indium (In or Ti or Al) was deposited at a low rate of 0.1 Å s$^{-1}$, followed by a deposition of a 70 nm-thick gold (Au) layer. The device was rinsed with acetone and isopropanol, subsequently to lift off. To ensure accurate active area definition, the top metal grid was precisely aligned with the bottom electrode. WSe$_2$ flakes extending beyond the grid were covered with metal to eliminate stray photocurrent contributions. A small additional photocurrent signal may arise from regions where WSe$_2$ bridges the bottom contact and the uncovered top metal mask, which was intentionally left open to prevent short-circuiting.

**Preparation and acquisition parameters for STEM specimens**

Cross-sectional STEM lamellas of our devices were prepared using a FEI Helios NanoLab G4 focused ion beam. STEM imaging was performed at 200 keV using a FEI Titan$^3$ G2 60-300 with a double-sided spherical aberration corrector. The probe convergence semi-angle was set to ~25 mrad, and ADF-STEM images were acquired over a 50–200 mrad range.

**Electrical characterization in dark**

Dark current-voltage (I-V) characteristics of the different devices were measured in a two-probe measurement system in an ambient atmosphere with a 4200-SCS Keithley.

**1−SUN I−V measurements.** To determine the PCE of the vertical cells under AM 1.5G, I–V measurements were performed using a solar simulator (Oriel Sol3A, Newport) connected to a Keithley 2440 source meter. The lamp intensity was calibrated with a Si reference cell (Newport, Oriel 91150 V) precisely placed at the sample position. I–V curves were recorded in a normal lab environment (temperature: 22 ºC; humidity: 35%). I−V characteristics were performed with a scan rate of 200 mV s$^{-1}$ and a dwell time of 30 ms. The external quantum efficiency was measured using a QuantX-300 Quantum Efficiency System (Newport). In this system, a monochromatic light beam was produced by passing the output of a 100 W xenon lamp through a monochromator. Since the device area was typically smaller than the illumination spot size, a 100 μm diameter pinhole was used, carefully aligned entirely onto the active area of the device. Measurements were taken at zero bias over a wavelength range of 300 to 1000 nm. However, our cells with similar individual layer thicknesses exhibit a variation in photovoltaic performance (see Table S1, Supporting Information). This refers to the variation in residues on surfaces/interfaces of different cells, which may influence the electrical/optoelectrical properties. The PCE, extracted from continuous maximum power point tracking, demonstrates the long-term operational stability of the hBN-encapsulated solar cell under ambient conditions. To monitor performance over time, I–V measurements were carried out continuously for 30 minutes per day and repeated periodically for 30 days.

**Photocurrent mapping**

Excitation was provided by a 405 nm continuous-wave laser (Coherent, CUBE 405-50C). The beam was attenuated with a mechanical slit and chopped at 630 Hz (Stanford Research Systems, SR540) before coupling into a polarization-maintaining single-mode optical fibre (NKT Photonics). The fibre output was collimated (Thorlabs, f = 50 mm, AC254-050-AB-ML), circularly polarized with a quarter-wave plate (Thorlabs, AQWP05M-600), and aligned with two steering mirrors before entering the microscope (Zeiss, Axioscope). Photocurrent mapping was performed by focusing the beam with a 100× objective (Nikon, TU Plan Fluor 100×/0.90) and scanning the sample using a piezo stage (PI, P-527). The photocurrent signal was amplified (Stanford Research Systems, SR570 low-noise current preamplifier) before detection with a lock-in amplifier (Stanford Research Systems, SR830). The chopped beam power was measured at 130 nW at the objective output, and photocurrent values were converted to EQE by measurement of a calibrated silicon detector (Enlitech, RC-S103011-E) under matched excitation conditions. Reflectance images required for local IQE calculations were collected in widefield imaging on the same microscope using epi-illumination from a 405 nm LED (Thorlabs, M405LP1). To achieve sufficient field of view, reflectance imaging was performed using a 40× objective (Nikon, Plan Fluor ELWD 40×/0.60) with detection on a sCMOS camera (Hamamatsu, ORCA-Flash4.0 LT3), calibrated by measurement of a uniform mirror with known reflectance (Semrock, MaxMirror). Photocurrent and reflectance images were aligned for IQE calculations by matching pixel sizes with spline interpolation and affine image registration in Python using the Advanced Normalization Tools (ANTs) package.[46]

**Acknowledgements**

MC and K.K.P. acknowledge funding from Horizon Europe UK Research and Innovation (UKRI) Underwrite MSCA (EP/Y028287/1). C.C. acknowledges the support of a Marshall Scholarship, Winton Scholarship and the Cambridge Trust. Z.L. acknowledges the financial support and Research Fellowship from the Herchel Smith Fund and King's College, Cambridge. L.L. acknowledges the support of Newton International Fellowship from the Royal Society (NIF\R1\242837). S.D.S. acknowledges the Royal Society and Tata Group (UF150033, URF\R\221026), the Leverhulme Trust (RPG-2021-191), and the EPSRC (EP/V012932/1) for funding. M.C. and Y.W. received funding from the European Research Council (ERC) Advanced Grant under the European Union's Horizon 2020 research and innovation programme (grant agreement GA 101019828-2D- LOTTO), EPSRC (EP/T026200/1, EP/T001038/1, EP/Z535680/1), Department of Science, Innovation and Technology and the Royal Academy of Engineering under the Chair in Emerging Technologies programme. This work was additionally supported by funding from the Henry Royce Institute (EP/R00661X/1) and Cambridge Royce facilities grant (EP/P024947/1). H.Y.J. acknowledges support from the National Research Foundation of Korea (NRF) grant funded by the Korea government (MSIT) (2022R1A2C2011109).

**Supplementary Information**

Description of charge transport in vertical diodes and ideality factor, low-magnification HAADF-STEM images of defective and ultraclean metal-$WSe_2$ interfaces, dark I-V characteristics of various Pt-$WSe_2$-In/Au devices, I-V curves of ultraclean vdW In/Au top contact diodes with different bottom contacts, atomic force micrograph (AFM) of Pt-$WSe_2$-In/Au device and line profiles, optical micrographs and corresponding $WSe_2$ thickness profiles in Pt-$WSe_2$-In/Au devices for various grid sizes, photocurrent density in Pt-$WSe_2$-In/Au devices with the variation of grid size, dark I-V characteristics of Pt-$WSe_2$-In/Au devices with varying top In/Au grid sizes and extracted series resistance, optical micrographs of Pt-$WSe_2$-In/Au devices with different grid sizes, variation of fill factor in Pt-$WSe_2$-In/Au devices with the $WSe_2$ film thickness, I-V characteristics of one batch of Pt-$WSe_2$-In/Au devices in 1-Sun and statistical distribution of PV parameters, I-V curve of a large-area device, I-V curves of a Pt-$WSe_2$-In/Au device in forward and reverse scan, hBN encapsulated device and stability in 1-Sun, EQE of Pt-$WSe_2$-In/Au devices with various grid sizes, reflectance and corresponding IQE map, AFM images of the Pt/$WSe_2$ surface at different magnifications with line profile, details of PV parameters for different Pt/$WSe_2$/In-Au devices under 1-SUN, and benchmark table.

# Supporting information

# High-efficiency WSe$_2$ photovoltaics enabled by ultra-clean van der Waals contacts

Kamal Kumar Paul[1]*, Cullen Chosy[2,3], Soumya Sarkar[1], Zhuangnan Li[1], Han Yan[1], Ye Wang[1], Leyi Loh[1], Lixin Liu[1], Hu Young Jeong[4], Samuel D. Stranks[2,3], Yan Wang[1], and Manish Chhowalla[1]*

[1]Department of Materials Science & Metallurgy, University of Cambridge, 27 Charles Babbage Road, Cambridge CB3 0FS, United Kingdom

[2]Department of Chemical Engineering and Biotechnology, University of Cambridge, Cambridge, United Kingdom

[3]Cavendish Laboratory, University of Cambridge, Cambridge, United Kingdom

[4]UNIST Central Research Facilities (UCRF) and School of Materials Science and Engineering, UNIST, Ulsan, South Korea

*kkp26@cam.ac.uk, mc209@cam.ac.uk

**Charge transport in vertical diodes and ideality factor**

A near-unity ideality factor is desired in a Schottky diode for ensuring the ultraclean metal-semiconductor interfaces, an un-pinned Fermi level free from interfacial trap states, leading to the negligible recombination current. The ideality factor (n) is determined by fitting the data to the Shockley diode equation in the dark:

$$I = I_S(e^{(qV/nkT)} - 1)$$

where $I_S$, q, k, and T are the reverse saturation current, elementary charge, Boltzmann constant, and temperature, respectively. Since defect-assisted recombination current usually dominates over diffusion current at low bias, the fitting was performed within the voltage range of 0-0.5 V. Defective Ti/Au top contact shows the poorest ideality factor of 2.38 (on/off ratio ~5.2×10$^3$), while Al/Au contact achieves a value of 2.06 (on/off ratio ~2.5×10$^4$). In contrast, the vdW top In/Au contact with Pd bottom contact shows a value of 1.12 (on/off ratio ~3.2×10$^8$), and the combination with a Pt bottom contact provides the best ideality factor of 1.1. The ideality factors are fully consistent with the reverse leakage current in various diodes. Therefore, a near-ideal Pt-WSe$_2$-In/Au Schottky diode with a record high on/off ratio (>1×10$^9$) promotes unidirectional current flow, efficient charge separation, and negligible recombination losses, making it a perfect candidate for superior photovoltaic applications (Fig. 2a). Under reverse bias, the defective contact diodes possess dominant leakage current through the midgap states (Fig. 2a), while vdW contact diodes with unpinned Fermi level achieve negligible midgap defect states, effectively reducing the leakage current (Fig. 2a). Under forward bias, the suppressed drain current in defective contact diodes is primarily due to the higher Schottky barrier height (Fig. 2b). In contrast, ultraclean, defect-free interface in vdW contact diodes enables efficient carrier transport due to low Schottky barrier and favorable band alignment, leading to near-unity ideality factor (Fig. 2b).[1]

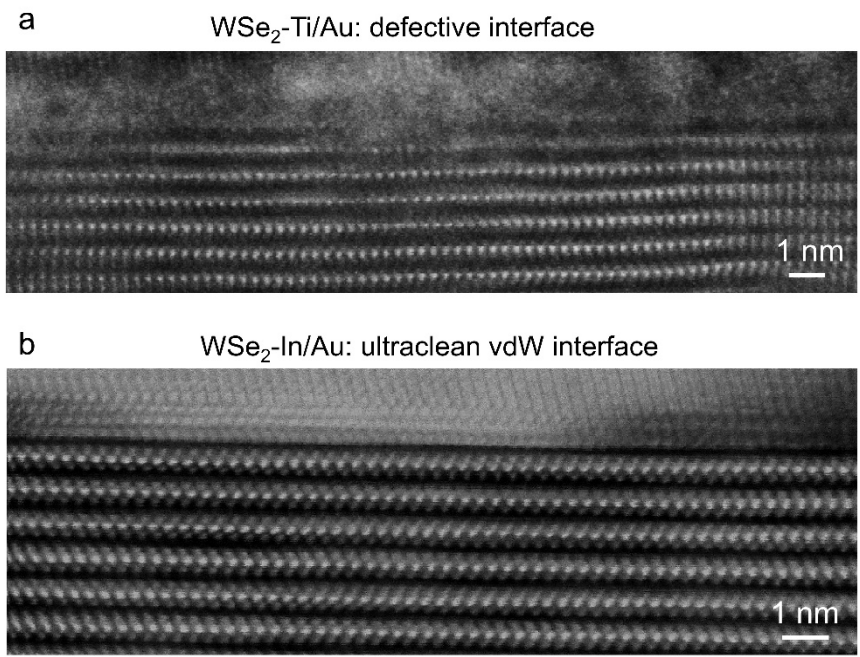

**Fig. S1:** Low-magnification HAADF-STEM images highlighting large-area interfacial contrast between (a) defective (Ti/Au or Al/Au) and (b) ultra-clean (In/Au) contacts, demonstrating widespread damage versus uniform, defect-free vdW contact regions.

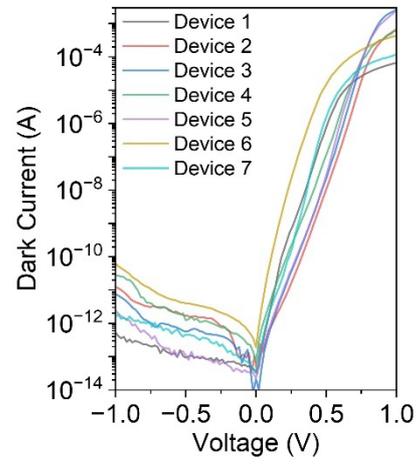

**Fig. S2:** Dark I-V characteristics of seven different Pt-WSe$_2$-In/Au devices, plotted on a semi-logarithmic scale.

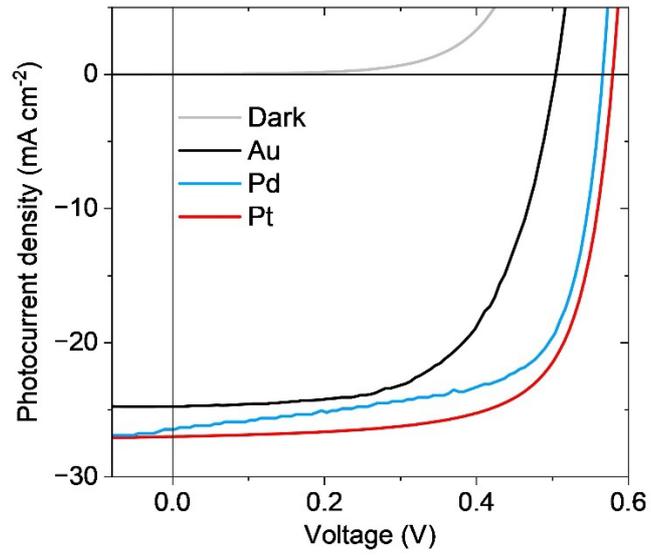

**Fig. S3:** I-V curves of ultraclean vdW In/Au top contact diodes with different bottom contacts (Au (black), Pd (blue), and Pt (red)).

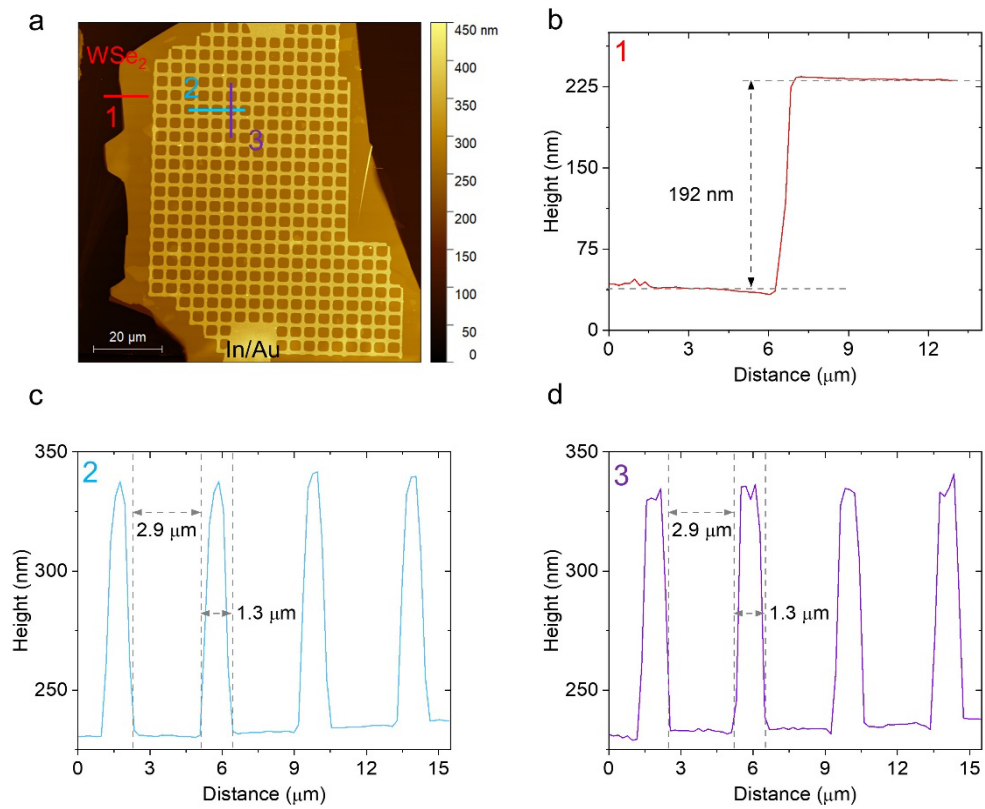

**Fig. S4: a,** Atomic force micrograph of a Pt-WSe$_2$-In/Au device, showing the top In/Au electrode with optical windows. **b,** The red line is the region from where the AFM height profile was collected, which shows the thickness of WSe$_2$ to be 192 nm. **c,** Horizontal grids along the blue line (**c**) and vertical grids along the violet line (**d**) show that the grid width is ~1.3 μm and the side length of the square optical window is 2.9 μm.

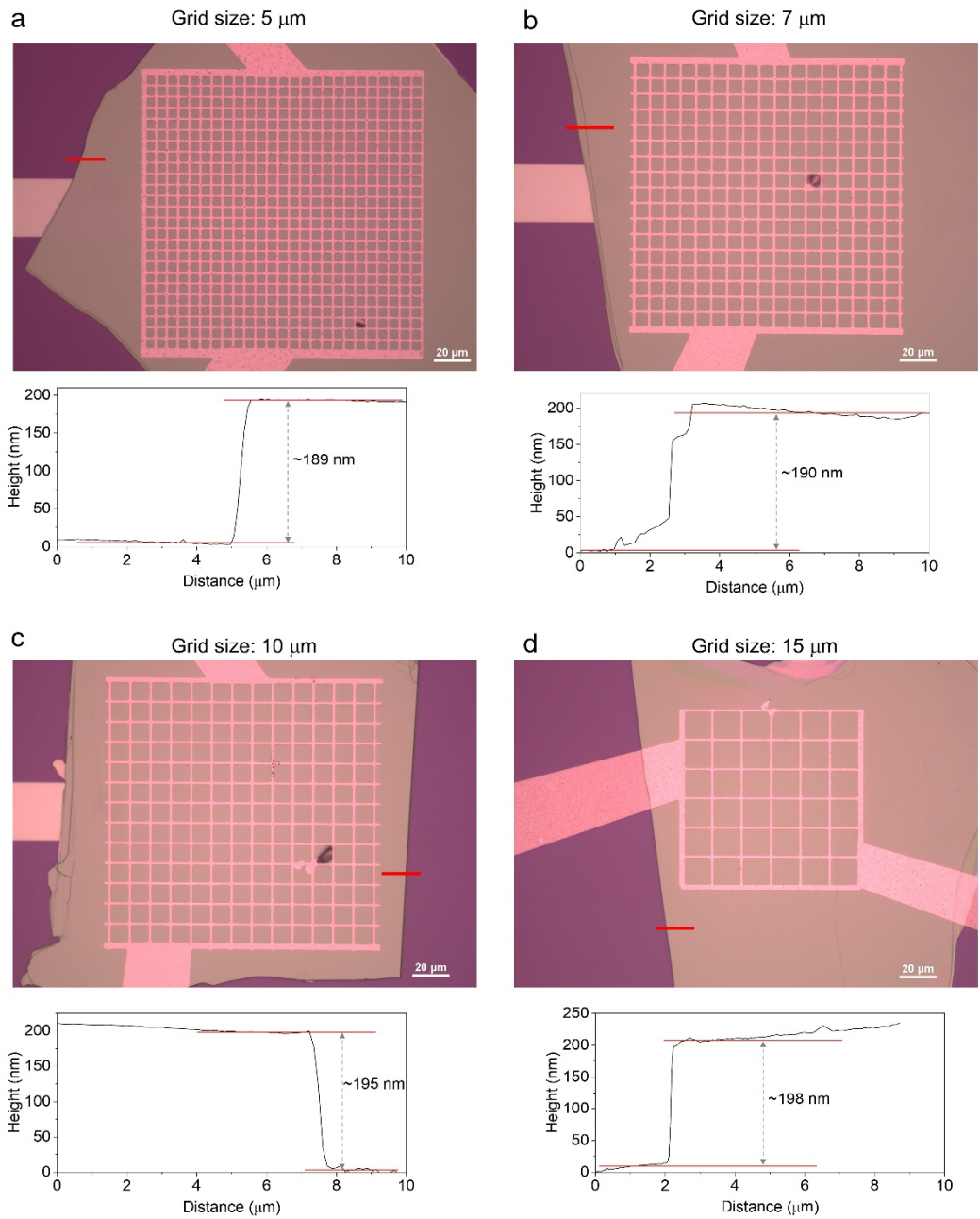

**Fig. S5:** Optical micrographs and the corresponding WSe$_2$ thickness profiles in Pt-WSe$_2$-In/Au devices for various grid sizes: **a,** 5 μm, **b,** 7 μm, **c,** 10 μm and **d,** 15 μm.

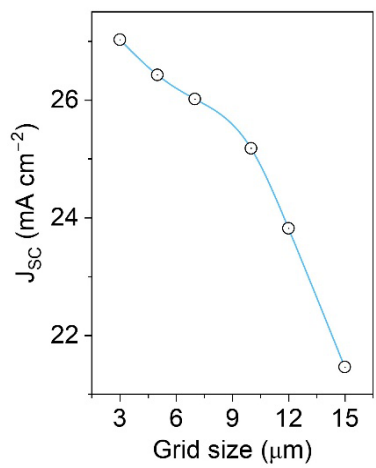

**Fig. S6:** Photocurrent density in Pt-WSe$_2$-In/Au devices with the variation of grid size.

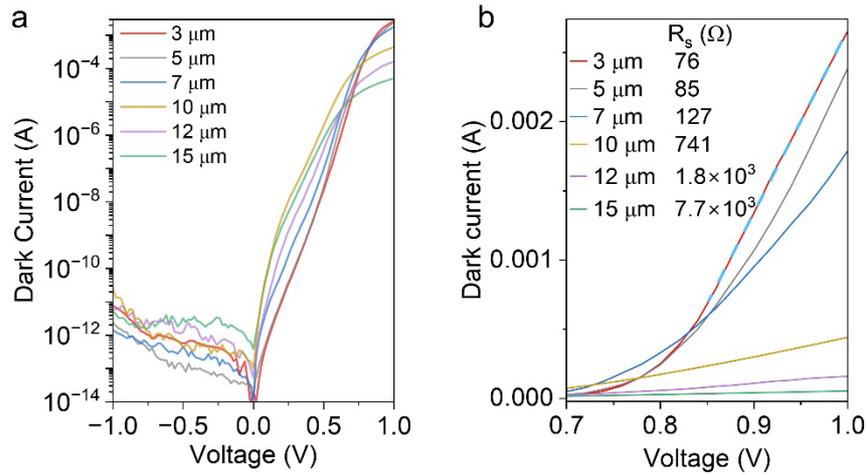

**Fig. S7: a,** Dark current–voltage characteristics of Pt–WSe$_2$–In/Au devices with varying top In/Au grid sizes. While reverse leakage remains largely unchanged, forward current decreases with increasing grid size, indicating a rise in series resistance. **b,** Forward bias I–V curves (0.7–1.0 V) plotted on a linear scale. Series resistance ($R_S$) is extracted from the slope of the linear region, as illustrated by the blue-dotted line. The inset summarizes the estimated series resistance values for each grid size. Devices with larger grids exhibit earlier forward current saturation and shallower slopes, reflecting higher series resistance, whereas smaller grid patterns enable steeper slopes with lower series resistance and more efficient carrier transport.

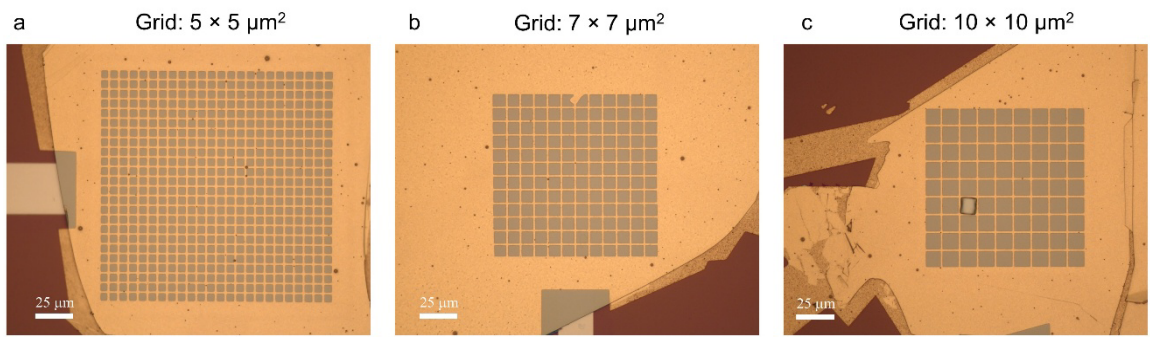

**Fig. S8:** Optical micrographs of Pt-WSe$_2$-In/Au devices with different grid sizes. (a–c): 5 × 5 µm$^2$, 7 × 7 µm$^2$, and 10 × 10 µm$^2$, respectively.

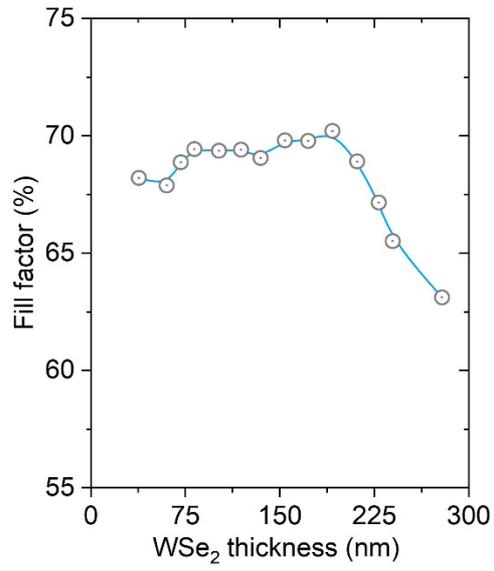

**Fig. S9:** Variation of fill factor in Pt-WSe$_2$-In/Au devices with the WSe$_2$ film thickness.

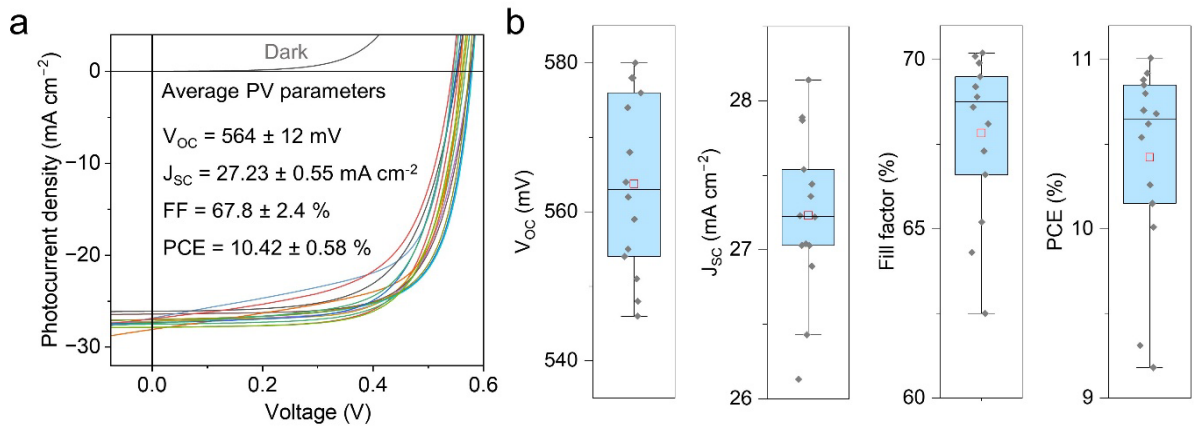

**Fig. S10: a,** The I-V characteristics of one batch (ten) of Pt-WSe$_2$-In/Au devices (5 × 5 μm$^2$ and 7 × 7 μm$^2$) under 1-Sun illumination (AM 1.5G). Inset shows the average values of various photovoltaic parameters with errors. **b,** Statistical distribution of $V_{OC}$, fill factor (FF) and PCE. The red box shows the corresponding mean value.

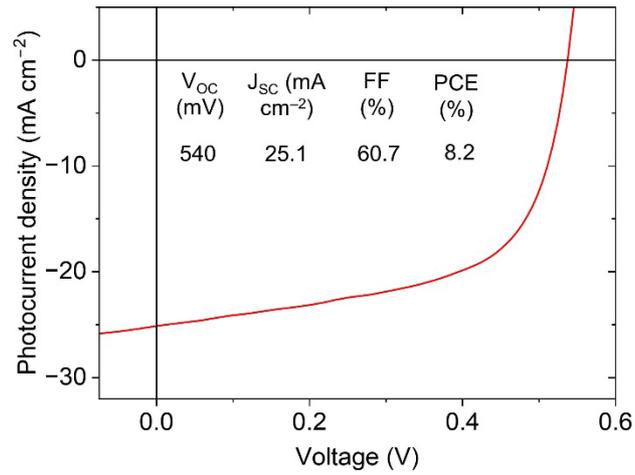

**Fig. S11:** The I-V curve of a large-area device with an active area of 0.32×0.32 mm². The device was fabricated by integrating nine WSe$_2$ flakes, ensuring a larger active region for enhanced photovoltaic performance. Inset shows the various photovoltaic parameters.

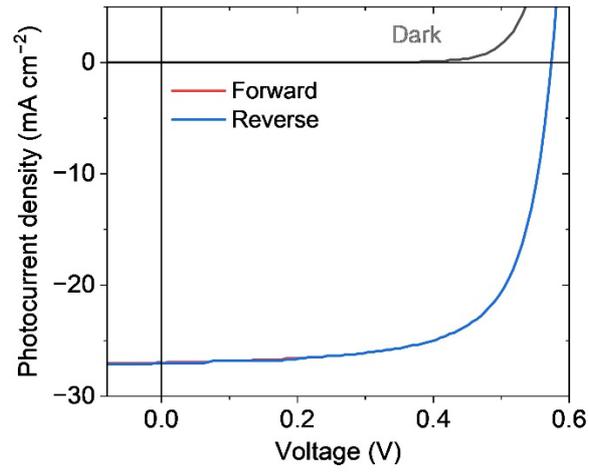

**Fig. S12:** I-V curves of a Pt-WSe$_2$-In/Au device under 1-Sun illumination in forward and reverse scan.

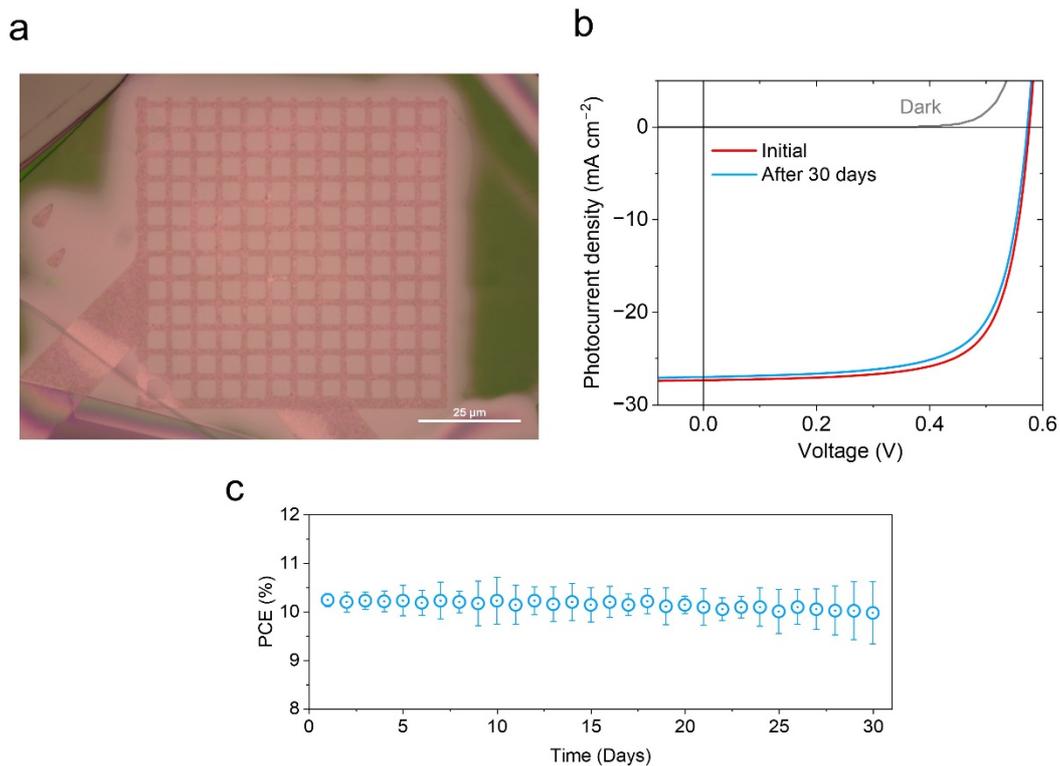

**Fig. S13: a,** Optical image of an hBN-encapsulated Pt-WSe$_2$-In/Au device. **b,** I-V curves of the device measured initially and after 30 days. The minimal change in the curves suggests strong stability of the hBN encapsulation. **c,** Operational stability of the device measured daily over 30 consecutive days, showing consistent performance with only slight variations, confirming the device's robustness under prolonged operation.

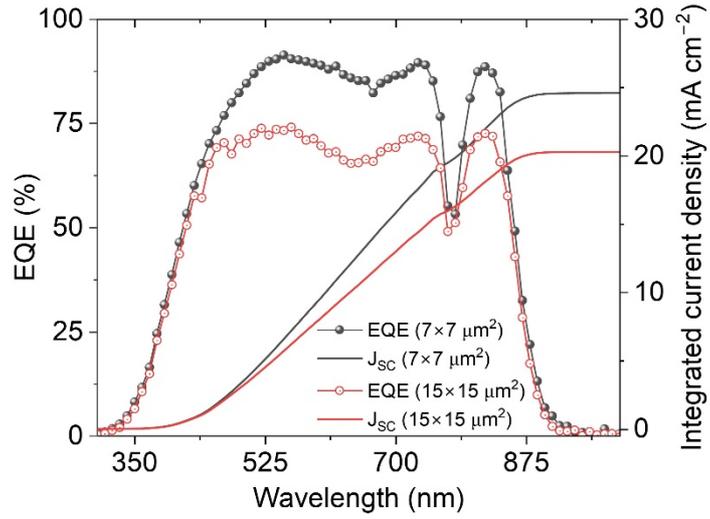

**Fig. S14:** EQE spectra of Pt–WSe$_2$–In/Au devices measured with grid sizes of 7 × 7 μm² and 15 × 15 μm² (left Y-axis). The corresponding integrated J$_{SC}$ plotted on the right Y-axis, estimated from EQE spectrum.

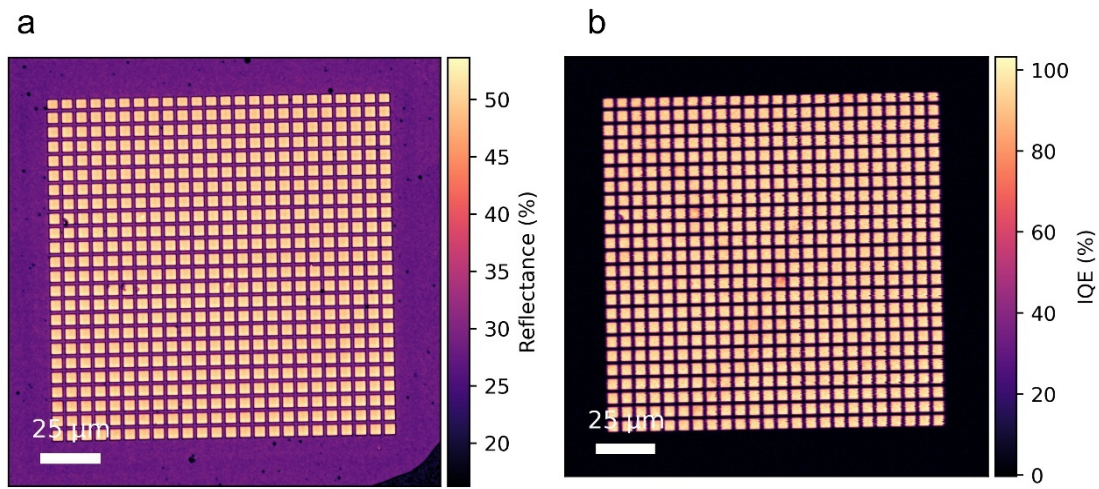

**Fig. S15: a,** Reflectance map of a Pt-WSe$_2$-In/Au device with grid size 5×5 μm$^2$ under a 405 nm laser. The colour contrast in the spatial map indicates uniform thickness across all optical windows. **b,** Corresponding IQE map considering the reflectance map shown in **a**.

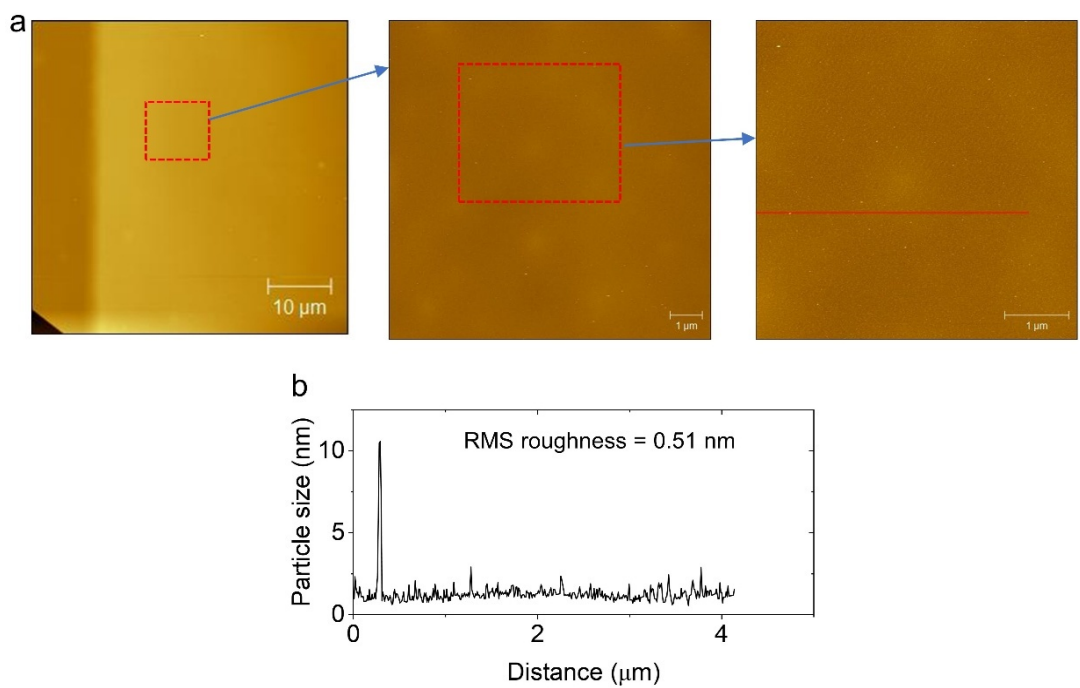

**Fig. S16: a,** Atomic force microscopy (AFM) images of the Pt/WSe$_2$ surface at different magnifications to evaluate surface cleanliness. **b,** Line profile reveals a smooth surface with minimal contamination or residues, indicating a clean and uniform Pt/WSe$_2$ interface.

**Table S1|** Details of photovoltaic parameters for different Pt/WSe$_2$/In-Au devices under 1-SUN illumination. Average parameters are also tabulated here. Standard error was calculated considering statistical analysis.

| Cell number | $V_{OC}$ (mV) | $J_{SC}$ (mA cm$^{-2}$) | FF (%) | PCE (%) |
|---|---|---|---|---|
| 1 | 555 | 27.87 | 69.2 | 10.7 |
| 2 | 578 | 27.04 | 69.9 | 10.92 |
| 3 | 568 | 27.54 | 68.9 | 10.8 |
| 4 | 576 | 27.22 | 68.1 | 10.68 |
| 5 | 564 | 27.23 | 68.6 | 10.54 |
| 6 | 580 | 27.03 | 70.2 | 11.01 |
| 7 | 578 | 26.43 | 69.5 | 10.62 |
| 8 | 574 | 27.03 | 70.1 | 10.88 |
| 9 | 559 | 28.14 | 65.2 | 10.26 |
| 10 | 546 | 26.89 | 62.5 | 9.18 |
| 11 | 562 | 27.89 | 69.2 | 10.85 |
| 12 | 554 | 26.13 | 64.3 | 9.31 |
| 13 | 551 | 27.36 | 67.3 | 10.15 |
| 14 | 548 | 27.44 | 66.6 | 10.01 |
| | **$V_{OC, average}$** **564 ± 12** | **$J_{SC, average}$** **27.23 ± 0.55** | **FF$_{, average}$** **67.8 ± 2.4** | **PCE$_{average}$** **10.42 ± 0.58** |

**Table S2|** Photovoltaic performance of different vertical vdW cells: $J_{SC}$ (short circuit current), $V_{OC}$ (open-circuit voltage), FF (fill factor), PCE (power conversion efficiency) under 1-Sun illumination.

| Device structure | $J_{SC}$ (mA cm$^{-2}$) | $V_{OC}$ (V) | FF | PCE (%) | Thickness (nm) | Ref |
|---|---|---|---|---|---|---|
| Gr/WS$_2$/Al | 16 | 0.7 | 0.29 | 3.3 | 37 | 2 |
| p-Gr/WSe$_2$/Au | 17.3 | 0.48 | 0.62 | 5.1 | 200 | 3 |
| Au/p-MoS$_2$/n-MoS$_2$/ITO | 20.9 | 0.28 | 0.47 | 2.8 | 120 | 4 |
| Cr-Pd/p-MoS$_2$/n-MoS$_2$/Cr-Pd-Cr | 3.0 | 0.6 | 0.22 | 0.4 | 11 | 5 |
| Au/WS$_2$/Al/ MoO$_X$ | 4.78 | 0.6 | 0.54 | 1.55 | 90 | 6 |
| Gr/WO$_X$/WSe$_2$/Pt | 19.61 | 0.47 | 0.59 | 5.44 | 242 | 7 |
| Pd/WSe$_2$/InO$_X$ | 20.78 | 0.483 | 0.64 | 6.37 | 100 | 8 |
| Pt/WS$_2$/Gr | 27.9 | 0.46 | 0.39 | 5 | 63 | 1 |
| Pt/WSe$_2$/ITO | 16.4 | 0.522 | 0.78 | 6.68 | 70 | 9 |
| Pt/WSe$_2$/Ti-Au | 11.9 | 0.202 | 0.465 | 1.12 | 150~250 | This work |
| Pt/WSe$_2$/Al-Au | 21.44 | 0.405 | 0.22 | 1.9 | 150~250 | This work |
| Au/WSe$_2$/In-Au | 24.75 | 0.505 | 0.614 | 7.68 | 150~250 | This work |
| Pd/WSe$_2$/In-Au | 26.45 | 0.567 | 0.673 | 10.1 | 150~250 | This work |
| **Pt/WSe$_2$/In-Au** | **27.03** | **0.580** | **0.702** | **11.01** | **150~250** | **This work** |